\documentclass[aps,prl,preprint,nopacs,superscriptaddress]{revtex4}

\usepackage{graphicx}
\usepackage{verbatim}
\usepackage{mathrsfs}
\usepackage{amsmath,amsfonts,amssymb}
\usepackage{graphicx}

\def\3{2.8in}    
\def\2{2.5in}
\def\4{3.0in}

\def \beq {\begin{equation}}
\def \eeq {\end{equation}}
\begin{document}

\title{Observation of a bulk 3D Dirac multiplet, Lifshitz transition, and nestled spin states in Na$_3$Bi}

\author{Su-Yang~Xu*}
\affiliation {Joseph Henry Laboratory, Department of Physics, Princeton University, Princeton, New Jersey 08544, USA}

\author{Chang~Liu*}
\affiliation {Joseph Henry Laboratory, Department of Physics, Princeton University, Princeton, New Jersey 08544, USA}

\author{S.~K.~Kushwaha}
\affiliation {Department of Chemistry, Princeton University, Princeton, New Jersey 08544, USA}

\author{T.-R.~Chang}
\affiliation{Department of Physics, National Tsing Hua University, Hsinchu 30013, Taiwan}

\author{J.~W.~Krizan}
\affiliation {Department of Chemistry, Princeton University, Princeton, New Jersey 08544, USA}

\author{R.~Sankar} \affiliation{Center for Condensed Matter Sciences, National Taiwan University, Taipei 10617, Taiwan}

\author{C.~M.~Polley}
\affiliation {MAX-lab, Lund University, S-22100 Lund, Sweden}

\author{J.~Adell}
\affiliation {MAX-lab, Lund University, S-22100 Lund, Sweden}

\author{T.~Balasubramanian}
\affiliation {MAX-lab, Lund University, S-22100 Lund, Sweden}

\author{K.~Miyamoto}
\affiliation {Hiroshima Synchrotron Radiation Center, Hiroshima University, 2-313 Kagamiyama, Higashi-Hiroshima 739-0046, Japan}

\author{N.~Alidoust}
\affiliation {Joseph Henry Laboratory, Department of Physics, Princeton University, Princeton, New Jersey 08544, USA}

\author{Guang~Bian}
\affiliation {Joseph Henry Laboratory, Department of Physics, Princeton University, Princeton, New Jersey 08544, USA}

\author{M.~Neupane}
\affiliation {Joseph Henry Laboratory, Department of Physics, Princeton University, Princeton, New Jersey 08544, USA}

\author{I.~Belopolski}
\affiliation {Joseph Henry Laboratory, Department of Physics, Princeton University, Princeton, New Jersey 08544, USA}

\author{H.-T.~Jeng}
\affiliation{Department of Physics, National Tsing Hua University, Hsinchu 30013, Taiwan}
\affiliation{Institute of Physics, Academia Sinica, Taipei 11529, Taiwan}

\author{C.-Y. Huang}
\affiliation{Department of Physics, National Sun Yat-Sen University, Kaohsiung 804, Taiwan}

\author{W.-F. Tsai}
\affiliation{Department of Physics, National Sun Yat-Sen University, Kaohsiung 804, Taiwan}

\author{H.~Lin}
\affiliation {Graphene Research Centre and Department of Physics, National University of Singapore 11754, Singapore}

\author{F.~C.~Chou} \affiliation{Center for Condensed Matter Sciences, National Taiwan University, Taipei 10617, Taiwan}

\author{T.~Okuda}
\affiliation {Hiroshima Synchrotron Radiation Center, Hiroshima University, 2-313 Kagamiyama, Higashi-Hiroshima 739-0046, Japan}

\author{A.~Bansil}
\affiliation {Department of Physics, Northeastern University, Boston, Massachusetts 02115, USA}

\author{R.~J.~Cava}
\affiliation {Department of Chemistry, Princeton University, Princeton, New Jersey 08544, USA}

\author{M.~Z.~Hasan}
\affiliation {Joseph Henry Laboratory, Department of Physics, Princeton University, Princeton, New Jersey 08544, USA}
\affiliation {Princeton Center for Complex Materials, Princeton University, Princeton, New Jersey 08544, USA}

\pacs{}

\date{\today}

\begin{abstract}
\textbf{Symmetry or topology protected Dirac fermion states in two and three dimensions constitute novel quantum systems that exhibit exotic physical phenomena \cite{Graphene, kimgeim, RMP, Zhang_RMP, Fang, Ashvin, Balent, 3D_Dirac, Dirac_3D, Dirac_semi, Dai, Yu Science QAH, Zhang Axion, Kim, Geim, Weyl superconductor}. However, none of the studied spin-orbit materials are suitable for realizing bulk multiplet Dirac states for the exploration of interacting Dirac physics. Here we present experimental evidence, for the first time, that the compound Na$_3$Bi hosts a bulk spin-orbit Dirac multiplet and their interaction or overlap leads to a Lifshitz transition in momentum space - a condition for realizing interactions involving Dirac states. By carefully preparing the samples at a non-natural-cleavage (100) crystalline surface, we uncover many novel electronic and spin properties in Na$_3$Bi by utilizing high resolution angle- and spin-resolved photoemission spectroscopy measurements. We observe two bulk 3D Dirac nodes that locate on the opposite sides of the bulk zone center point $\Gamma$, which exhibit a Fermi surface Lifshitz transition and a saddle point singularity. Furthermore, our data shows evidence for the possible existence of theoretically predicted weak 2D nontrivial spin-orbit surface state with helical spin polarization that are nestled between the two bulk Dirac cones, consistent with the theoretically calculated (100) surface-arc-modes. Our main experimental observation of a rich multiplet of Dirac structure and the Lifshitz transition opens the door for inducing electronic instabilities and correlated physical phenomena in Na$_3$Bi, and paves the way for the engineering of novel topological states using Na$_3$Bi predicted in recent theory \cite{Yu Science QAH, Zhang Axion, Kim, Geim, Fang, Ashvin, Balent, 3D_Dirac, Dirac_3D, Dirac_semi, Dai}.}
\end{abstract}

\maketitle

\section{Introduction}

With the explosion of research interest on 2D Dirac electronic states realized in graphene and surfaces of 3D topological insulators (TI) \cite{Graphene, kimgeim,RMP, Zhang_RMP}, a new research front that focus on the Dirac electronic phases of matter in three dimensions has emerged. In a 3D bulk Dirac semimetal phase, the bulk conduction and valence bands only touch at discrete locations in a 3D bulk Brillouin zone (BZ) and disperse nearly linearly along all three momentum space directions, making it a natural hypercone analog of graphene. Although such materials are known previously, recent interest is actually focused on spin-orbit based 3D bulk Dirac semimetal phase since the spin-orbit coupling can drive exotic topological phenomena and quantum transport in such materials as the Weyl phases, high temperature linear quantum magnetoresistance and topological magnetic phases \cite{Fang, Ashvin, Balent, 3D_Dirac, Dirac_3D, Dirac_semi, Dai}. Spin-orbit based 3D Dirac materials are actually quite rare. The strongly spin-orbit coupled 3D bulk Dirac semimetal phase is predicted \cite{3D_Dirac} and experimentally achieved in the topological phase transition (TPT) between a conventional insulator and a 3D topological insulator, such as in BiTl(S$_{1-\delta}$Se$_{\delta}$)$_2$ and (Bi$_{1-\delta}$In$_\delta$)$_2$Se$_3$ systems \cite{Suyang, Oh}. However, fine-tuning of the chemical doping/alloying composition is difficult to control, which is very sensitive to chemical synthesis conditions. More importantly, in this approach, the 3D bulk Dirac cone and the 2D nontrivial surface states (SSs) cannot coexist, because the 3D bulk Dirac cone only occurs at the critical point of the TPT, whereas the 2D nontrivial surface states only occur at the inverted side of the transition \cite{3D_Dirac}.

Recent theoretical advances have provided another, more exotic, route to realizing the spin-orbit 3D Dirac semimetal via additional symmetry protection \cite{Dirac_3D, Dirac_semi, Dai}. In this scenario, the system usually has an inverted spin-orbit bulk band structure similar to a TI. However, interestingly, spin-orbit coupling cannot open up a bulk energy gap at some special momentum-space locations in the bulk BZ due to the protection ensured by certain crystalline space group symmetries \cite{Dirac_3D}, and the bulk Dirac band touching is therefore preserved without the requirement of fine tuning to the TPT critical point. Following the theoretical proposals, photoemission experiments have studied the (001)* \cite{Note} surface electronic structure in Cd$_3$As$_2$ (see, preprints \cite{CdAs_Hasan, CdAs_Cava}) and (see, preprints Na$_3$Bi \cite{Chen_Na3Bi}). Although in both materials a \textit{single} bulk Dirac cone locating at the surface BZ center has been identified \cite{CdAs_Hasan, CdAs_Cava, Chen_Na3Bi}, the most exciting physics in these systems, including the presence of bulk Dirac multiplet (multiple bulk Dirac cones), their Lifshitz transition and saddle point singularities, as well as the potential spin-momentum locked nontrivial surface (2D) state arcs that are predicted to nestle between the two bulk Dirac cones \cite{Dirac_semi}, have not been experimentally observed and studied. In this article, we experimentally discover these novel properties in spin-orbit bulk 3D Dirac material Na$_3$Bi by systematically studying the electronic structure at a carefully prepared non-natural cleavage surface (100). Using angle-resolved photoemission spectroscopy (ARPES), we observe two bulk 3D Dirac nodes on the opposite sides of the bulk BZ center. Our data shows that these two bulk Dirac cones exhibit a Fermi surface Lifshitz transition and a saddle point singularity, opening the door for inducing electronic instability and correlated physics phenomena in bulk Dirac semimetals. We further show some evidence for the possible existence of theoretically predicted weak 2D nontrivial spin-orbit surface state with helical spin polarization that is nestled between the two bulk Dirac cones, consistent with the theoretically calculated (100) surface-arc-modes. Our experimental results identify the first real example of a novel Dirac-cone-multiplet electronic structure featuring saddle point singularity, with additional signature for nontrivial surface states, which provides a fertile and rich playground for many exotic topological states and phenomena, such as quantum anomalous Hall effect \cite{Fang, Yu Science QAH}, axion electrodynamics \cite{Zhang Axion}, Weyl topological charges \cite{Ashvin, Balent, Dirac_semi, Dai} and superconductors \cite{Weyl superconductor} to be engineered.

\section{Results}

\begin{figure*}
\centering
\includegraphics[width=15cm]{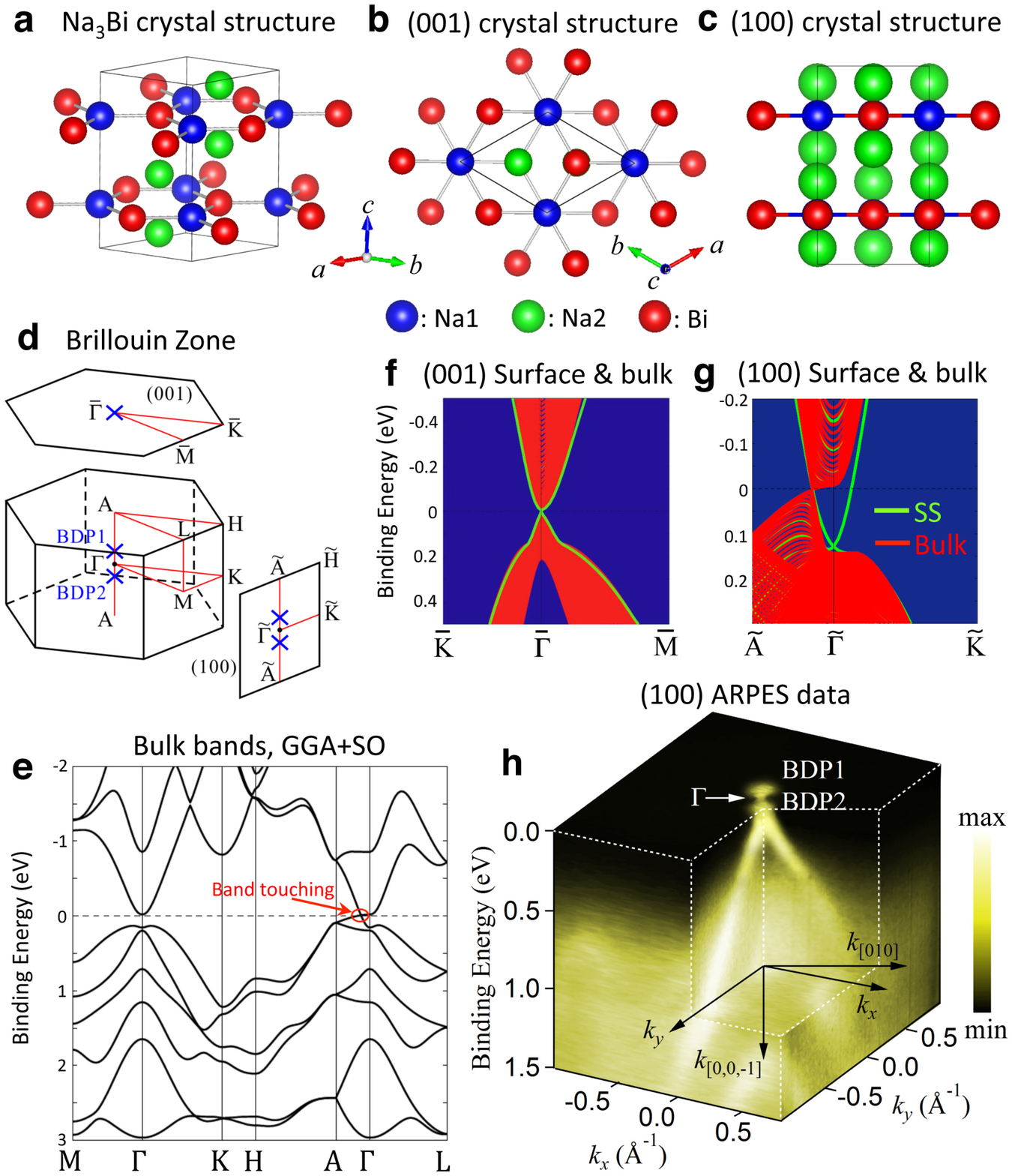}
\caption{\textbf{Characterization of Na$_3$Bi system.} \textbf{a}, Crystal structure of Na$_3$Bi. The two Na sites and Bi atoms are marked with different colors. \textbf{b}-\textbf{c}, Projected crystal structure at the (001) and (100) surfaces. \textbf{d}, Structure of bulk and surface Brillouin zone at (001) and (100) surfaces. Bulk Dirac nodes are marked by blue crosses. Note that the two cones project to the same $\bar{\Gamma}$ point on the (001) surface, while they are separated in momentum when studied at the (100) surface. \textbf{e}, First principles bulk band calculation for Na$_3$Bi. It is clear from the calculation that the band touching happens along the $A$-$\Gamma$-$A$ direction close to the zone center.}\label{Characterization}
\end{figure*}

Na$_3$Bi is a semimetal that crystalizes in the hexagonal $P6_3/mmc$ crystal structure with $a=5.448$ $\textrm{\AA}$ and $c=9.655$ $\textrm{\AA}$ \cite{Na3Bi_crystal} ($a^{-1}/c^{-1}\simeq1.8$), as shown in Fig. \ref{Characterization}\textbf{a}. It is a layered structure consisting of a honeycomb atomic plane where the Na and Bi atoms are situated on the A and B sites respectively, sandwiched by two layers of Na atoms arranged in a triangular fashion (Figs. \ref{Characterization}\textbf{b}-\textbf{c}). The single crystals are extremely air-sensitive and are found to degrade in air within seconds. Our first principles bulk band calculations (Fig. \ref{Characterization}\textbf{e}) show that its lowest bulk conduction and valence bands are composed of Bi $6p_{x,y,z}$ and Na $3s$ orbitals \cite{Dirac_semi}. These two bands possess a bulk band inversion of about $\sim0.3$ eV at the bulk BZ center $\Gamma$ \cite{Dirac_semi}. Due to the protection of an additional three-fold rotational symmetry along the [001] crystalline direction, two bulk Dirac nodes are predicted to exist as shown by the blue crosses in Fig. \ref{Characterization}\textbf{d}. However, it is important to notice that these two bulk Dirac nodes locate along the $A$-$\Gamma$-$A$ [001] direction. Therefore, if one study the (001) surface electronic structure, as done in previous experiments \cite{Chen_Na3Bi}, the two bulk Dirac nodes, as well as the possible nontrivial spin-orbit surface states (due to bulk inversion at $\Gamma$) will all be projected to the same location in in-plane momentum (the surface BZ center $\bar{\Gamma}$) and energy (Fermi level $E_F$) space, as shown by the (001) slab calculation in Fig. \ref{Characterization}\textbf{f}, making it difficult  to separate, isolate, and systematically study them via spectroscopic methods. We overcome this obstacle by studying another crystalline surface, namely the (100) plane. As shown in Fig. \ref{Characterization}\textbf{g}, along the [100] direction, the two bulk Dirac cones and the surface states are projected onto different momentum-space locations, thus making it possible to study this exotic electronic structure (here we followed Ref. \cite{Dirac_semi} and call the surface states in Na$_3$Bi as nontrivial surface state arcs rather than topological surface states, because a topological Z$_2$ order strictly requires a full bulk energy gap). In light of the above arguments, we systematically study the (100) electronic structure of Na$_3$Bi. The three dimensional band dispersion presented in Fig. \ref{Characterization}\textbf{h} outlines our experimental results. Even at the first glance one recognizes two important features. First, the two Dirac nodes that are separated in the momentum space are clearly visible (see also Fig. \ref{3Dcones}\textbf{a}). Second, the dispersive pattern of one of the cones is linear along both $k_x$ and $k_y$, which unambiguously establishes its Dirac semimetallic nature. These results endorse the critical importance of studying the (100) surface instead of the (001) natural cleavage plane in Na$_3$Bi.

Fig. \ref{3Dcones} shows the energy-momentum dispersion of the two bulk Dirac cones in the vicinity of the $\Gamma$ point. One should keep in mind that it is not easy to achieve a well defined band dispersion pattern due to technical difficulties in preparing these extremely air-sensitive samples. Fig. \ref{3Dcones}\textbf{a} presents the ARPES constant energy maps. The length ratio between the long and short edge of the surface BZ (yellow boxes) is measured to be about $1.8 \simeq a^{-1}/c^{-1}$, consistent with a (100) cleaving surface. We observe rare rectangular-like Fermi surfaces in the rectangular-shaped surface BZ at the Na$_3$Bi (100) surface, making the studies on this side surface more interesting. At the Fermi level, the ARPES intensity is mostly located at two $k$-space locations, equally spaced but on opposite sides of the zone center. A slight increase of binding energy up to 50 meV results in enlargement of the Fermi nodes, indicating the $\Lambda$-shaped Dirac cone dispersion. At higher binding energies this two cones merge into a single band contour, which then further expands to form a neck region around the $\tilde{A}$ points (top of the 3D BZ). In other words, Fig. \ref{3Dcones}\textbf{a} reveals two separated bulk Dirac cones whose band crossing appears right at the Fermi level. Fig. \ref{3Dcones}\textbf{b} presents several typical ARPES $k$-$E$ maps along various cutting directions (black solid lines in the left panel of Fig. \ref{3Dcones}\textbf{b}). A clear linear dispersion is seen for the bulk Dirac cones in all of these maps, establishing the existence of massless bulk quasiparticles. From Cuts 1 and 2 we determine the Fermi velocity along $k_{[010]}$ to be $v_F^{[010]} \sim 1.5$ eV$\cdot\textrm{\AA}=2.3\times10^5$ m/s, comparable to the in-plane Fermi velocities of the Na$_3$Bi Dirac cone measured at the (001) surface \cite{Chen_Na3Bi}. The asymmetric cone shape seen from Cut 3 at higher binding energies is due to the fact that the right cone reaches the Fermi level at this cutting location, while the left cone only evolves to a finite binding energy before $E_F$.

In Fig. \ref{kz} we study the out-of-plane [100] band dispersion along the third $k$-space direction ($k_z$) by varying the incident photon energies of the synchrotron radiation. Fig. \ref{kz}\textbf{a} shows ARPES constant energy maps covering a $k$-space region of $-0.4$ $\textrm{\AA}^{-1}$ $< k_y, k_z < 0.4$ $\textrm{\AA}^{-1}$, which corresponds to a photon energy range of $40$ eV $< h\nu < 76$ eV. As illustrated in Fig. \ref{kz}\textbf{b}, because the sample is cleaved \textit{in situ} on the (100) surface, $k_z$ corresponds to the [100] direction in the $k$-space, which is parallel to the $a$ crystallographic direction. From both band calculations and previous ARPES results, we have $v_F^{[100]} \sim v_F^{[010]}$. Thus we expect the constant energy maps to demonstrate an enlarging, circular-like band contour. From Fig. \ref{kz}\textbf{a} one realizes that this is indeed the case. The single intensity dot seen at $E_F$ around $h\nu = 58$ eV gradually expands to a circular contour, and remains circular-like up to a binding energy of 600 meV. In the $E$-$k_z$ ($E$-$k_{[100]}$) cut shown in Fig. \ref{kz}\textbf{c}, a clear dispersion along the out-of-plane direction (consistent with linearity) is observed. However, the linearity of the band along the $k_z$ [100] direction cannot be strictly concluded in Fig. \ref{kz}\textbf{c} due to the limited resolution along $k_z$ axis (see discussion in next paragraph). We overcome this difficulty by also studying the (001) surface (see Fig. 5). In Fig. \ref{kz}\textbf{d} the same fact is emphasized again by showing the $k_y$-$E$ maps for multiple photon energies. One realizes from Fig. \ref{kz}\textbf{d} that the Dirac band touches the Fermi level only at $h\nu = 58$ eV. At lower or higher photon energies the band recede linearly to higher binding energies. Therefore, the 3D Dirac point appears at $h\nu = 58$ eV. Combining the results from Figs. \ref{3Dcones} and \ref{kz}, we conclude that the electronic structure of Na$_3$Bi features a \textit{pair} of 3D bulk Dirac cones along the $\Gamma$-$A$ direction, whose $k_{[001]}$ momenta differ by $\sim 0.2$ $\textrm{\AA}^{-1}$.

Since the two bulk Dirac cones are located very close to one another in the momentum space, they inevitably touch and hybridize with each other, giving rise to a topological change in the band contours, also known as a Lifshitz transition in the electronic structure. Because the two cones locate at the $A$-$\Gamma$-$A$ axis, one needs to view the bands from a side surface of the crystal, e.g. the (100) surface, in order to reveal such intriguing band evolution. With samples prepared on the (001) surface, detailed study of this novel property would be extremely challenging, if not impossible, since in that surface both cones project onto the zone center without a $k_{\parallel}$ separation \cite{Chen_Na3Bi}. Fig. \ref{Lifshitz} demonstrates our systematic ARPES investigation on the Lifshitz transition as well as the associated saddle point band structure. Fig. \ref{Lifshitz}\textbf{a} presents a set of ARPES constant energy maps with higher momentum resolution than the one shown in Fig. \ref{3Dcones}. Besides the two bulk Dirac cones presented already in Fig. \ref{3Dcones}, we emphasize here the change of band structure when the two cones touch and merge into a single contour. Experimentally, this Lifshitz transition happens at a binding energy of $\sim144$ meV, at which the two holelike bands transform to a diamond-shaped loop. At higher binding energies, this loop enlarges while another concentric contour starts to appear at the zone center. This trend of band evolution is nicely reproduced in first principles band calculations, as shown in Fig. \ref{Lifshitz}\textbf{b} where a set of 3D ($k_{[100]}$, $k_{[010]}$, $k_{[001]}$) constant energy maps at selected binding energies are vividly presented. The theoretical critical point of the Lifshitz transition is $E_B = 134$ meV (only 10 meV away from the experimentally observed value). It can be seen  in Fig. \ref{Lifshitz}\textbf{a} that the two cones touch at two special momenta, ($k_x$, $k_y$) = ($\pm0.1$, 0) $\textrm{\AA}^{-1}$. To directly identify the saddle point band structure as a result of the Lifshitz transition in experiments, we center our ARPES detector at one of the special momenta at ($k_x$, $k_y$) = ($+0.1$, 0) $\textrm{\AA}^{-1}$ and study two $E$-$k$ cuts, namely, Cuts X and Y as defined in the $3^{\mathrm{rd}}$ panel in Fig. \ref{Lifshitz}\textbf{a}. As shown in Fig. \ref{Lifshitz}\textbf{d}, the hybridized band reaches its local energy maximum at ($k_x$, $k_y$) = ($+0.1$, 0) $\textrm{\AA}^{-1}$ along $k_x$, while arriving a local energy minimum along $k_y$. This behavior defines a saddle point singularity of band structure. At Fig. \ref{Lifshitz}\textbf{e}, an anomalous increase of local density of state (DOS) is observed by integrating the ARPES intensity around the saddle point, consistent with local DOS maximum in saddle point type of van Hove singularity in three-dimensions. We emphasize here that our observation of bulk Dirac multiplet, their Lifshitz transition and saddle point singularity is made possible thanks to our choice of the (100) cleavage surface. Although the two cone structure is theoretically visible also from the (001) termination (in this case we scan $k_z$ to view the two cones), the $k_{\bot}$ resolution in ARPES is usually substantially lower than that along the in-plane directions. This is because in ARPES, the in-plane $k$-space is mapped by only changing the sample normal angle with respect to the detector, whereas the out-of-plane $k$-direction is scanned by varying photon energy, which usually changes other conditions such as the photoemission cross-section of each band and the photoemission matrix element (e.g. compare Fig. \ref{3Dcones}\textbf{b} ($k_{\|}$) with Fig. \ref{kz}\textbf{c} ($k_{\perp}$)). These effects render the observation of the two separated cones very difficult (these cones are separated by a mere 0.2 $\textrm{\AA}^{-1}$ apart), and hinder the in depth study of the Lifshitz transition and saddle point singularity. In Fig. \ref{Lifshitz}\textbf{c} we compare and contrast the Lifshitz transition observed here with that observed in the surface states of a topological crystalline insulator Pb$_{0.7}$Sn$_{0.3}$Se. In the $n$-typed sample of Pb$_{0.7}$Sn$_{0.3}$Se, the two \textit{surface} Dirac cones hybridizes at $E_B = 23$ and 96 meV, signified also by saddle point singularities that are reported from ARPES as well as STM studies. The critical difference between these two cases is that for Na$_3$Bi, \textit{bulk} Dirac quasiparticles, instead of the \textit{surface} Dirac electron gas in the case of TCI \cite{Okada}, merge and hybridize.


Having established the existence of a pair of bulk Dirac cones in Na$_3$Bi, in Fig. \ref{TSS} we search for signatures of possible nontrivial surface states inspired by the theoretical prediction of the 0.3 eV bulk band inversion at $\Gamma$. In Fig. \ref{TSS}\textbf{a} we show again the Fermi surface of Na$_3$Bi at the (100) surface termination. Besides the two bulk Dirac nodes we discussed in length before, the most important observation here is the extra ARPES intensity in between the two nodes (i.e., right at the $\tilde{\Gamma}$ point). This curious intensity are found to be ``nestled'' between the two bulk Dirac cones, like the handle of a dumbbell, and importantly, it does not show up in our bulk calculations (Fig. \ref{Characterization}\textbf{e}) and is in agreement with the theoretical Fermi surface from our slab calculations (Fig. \ref{TSS}\textbf{b}). This state is even more clearly visualized at $E_B = 100$ meV, at which the bulk bands are observed to be two unconnected pockets (this is naturally so because $E_B = 100$ meV is above the Lifshitz transition at 144 meV). Furthermore, two additional curves (guided by the green dotted line) are observed to connect the two bulk pockets, which is consistent with the nontrivial surface state Fermi arcs predicted in Ref. \cite{Dirac_semi}. Better resolving the detailed electronic structure of the surface Fermi arcs will require $n-$doping to the surface, in order to study the constant energy contour at an energy well-above the surface Dirac point (surface upper Dirac band) so that the surface states span in a wider range in the $k$-space. Our data support the existence of surface Fermi arcs in Na$_3$Bi, which indicates exotic surface electrical and spin transport behaviors. To further investigate the origin of this $\tilde{\Gamma}$ electronic state, in Fig. \ref{TSS}\textbf{c} we present three typical ARPES $k$-$E$ maps cutting across the tip of the two bulk cones (Cuts 1 and 3) and the $\tilde{\Gamma}$ point (Cut 2). We observe in Cuts 1 and 3 the linear dispersion of the two bulk Dirac cones whose electronic states extend to the Fermi level as the Dirac points. In Cut 2, if there was no SS, one would not expect any electronic state that crosses the Fermi level. However, we observe here a single propeller-like band centered at the $\tilde{\Gamma}$ point. In a duplicated map of Cut 2 (Fig. \ref{TSS}\textbf{e}) the ARPES result is overlayed with lines that indicate the bulk and surface as guides to the eye. From this panel we see that the bulk bands are M-shaped and comes to a local minimum at the $\tilde{\Gamma}$ point. This is a signature of spin-orbit hybridization as well as inversion of the conduction and valence bands \cite{Suyang}. Furthermore, it is clear from this panel that the small $\tilde{\Gamma}$ state does not correspond to any of the bulk bands. This result is nicely reproduced in our slab calculations along the same directions as Cut 1 and Cut 2 (Fig. \ref{TSS}\textbf{d}). These observations serve as strong evidence that the $\tilde{\Gamma}$ state is a surface-arc electronic structure and is a result of the single bulk band inversion at $\Gamma$. ARPES results are in good agreement with the theoretical calculation along these three cuts.

In light of the above argument that the small $\tilde{\Gamma}$ state should be a SS, we wish to study the texture of its spin polarization, similar to what we previously presented for the well studied topological insulators \cite{RMP}. Since the SS exists in a very small range in energy and momentum, it is hard to resolve the $k$-space anisotropy of its spin polarization at the (100) surface, considering the limited $k$ and $E$ resolution of spin-resolved ARPES. Instead, we solve this problem by measuring its spin polarization at (001), the natural cleavage plane of Na$_3$Bi. Along the (001) surface, the two bulk Dirac cones, as well as the surface Dirac cone, are expected to all project onto the same $\bar{\Gamma}$ point, and their Dirac crossings are predicted to be degenerated at the Fermi energy. In other words, the SSs are expected to disperse along the \textit{outer boundary} of the 3D bulk Dirac cone energy-momentum continuum. Figs. \ref{Characterization}\textbf{f,g} show the theoretically calculated single surface state at surface BZ center $\bar{\Gamma}$ ($\tilde{\Gamma}$) at the (001) and (100) surfaces, both of which are a result of the same bulk band inversion at the bulk BZ center $\Gamma$. The left panel of Fig. \ref{TSS}\textbf{f} shows the spin integrated ARPES dispersion at (001) surface along the [100] direction, in which we indeed see a \textit{linearly}-dispersive cone feature with the Dirac point at the Fermi level. Besides the spin polarization, one important experimental conclusion is that we now also establish that the linear dispersion nature along the [100] direction, which therefore conclusively identifies the 3D Dirac nature of the bulk cones in Na$_3$Bi. This is uniquely enabled by our careful method of studying both the (100) and (001) surfaces. Now we present the results of our spin polarization measurements at the (001) surface. Two momenta on the opposite sides of the Dirac node at $k_1=-0.1$ $\textrm{\AA}^{-1}$ and $k_2=+0.02$ $\textrm{\AA}^{-1}$ are selected (see dotted lines in Fig. \ref{TSS}\textbf{f}) for spin-resolved measurements. In the two right panels of Fig. \ref{TSS}\textbf{f}, clear imbalance of tangential spin polarization is detected on opposite sides of the $\bar{\Gamma}$ point, with a maximum of $\sim15\%$ spin polarization, which is smaller than that observed in the SSs of a regular TI such as Bi$_2$Se$_3$. This is intuitive because the SSs overlap with the bulk 3D cone in $k$-space at the (001) surface, and therefore its spin polarization is reduced because they are surface resonance states \cite{Helical_metal}. The helical spin texture observed here provide strong evidence for the nontrivial nature of the surface states we observed in Na$_3$Bi, consistent with the predicted band inversion at the bulk BZ center. In summary, in Fig. \ref{TSS} we provide two pieces of strong experimental evidence for the existence of 2D nontrivial surface states in the surfaces of Na$_3$Bi. These evidence are the observation of (1) a low energy electronic state consistent with the theoretically calculated nontrivial surface state Fermi arc electronic structure, and (2) helical spin polarization at another surface termination but resulted from the same bulk band inversion at the bulk BZ center $\Gamma$ (same topological origin). Both evidence demonstrate the existence of nontrivial surface states as a result of the predicted band inversion at $\Gamma$.


\section{Discussion}

We discuss the emergent topological and quantum phenomena enabled by our experimental discovery. Saddle point singularities in 2D Dirac electron gas has been extensively studied. For example, striking Hofstadter's butterfly spectrum and fractional quantum Hall effect have been experimentally observed in twisted graphene systems due to the saddle point singularity in its Dirac band structure \cite{Kim, Geim}. Moreover, formation of multiple 2D Dirac nodes has been identified in the surface states of topological crystalline insulators (Fig. \ref{Lifshitz}\textbf{c}) \cite{Okada}, which enriches such physics in 2D by adding a topological interplay. In three-dimensions, Lifshitz transitions have been observed in high-$T_c$ copper and iron based superconductors \cite{BSCCO, Pnictide}, which are believed to be the keys to giving rise to their unconventional superconductivity \cite{BSCCO, Pnictide, Quasi-2D}. However, these superconductors lack strong spin-orbit coupling, which is critical for realization of topological physics. Here, our observation of bulk Dirac multiplet realizes the first Lifshitz transition and saddle point singularity in a strongly \textit{spin-orbit} coupled \textit{3D Dirac} material. For van Hove singularity (VHS) in three-dimensions, the DOS itself is usually not divergent (although its derivative is). A divergence in DOS can be reached if a system or a particular low energy band is quasi-2D, like the case in the VHS in high-$T_c$ superconductors \cite{Quasi-2D, BSCCO}. In Na$_3$Bi, the magnitude of the DOS peak at Lifshitz transition energy can be effectively tuned by changing interlayer coupling strength, possibly reaching divergence in the very weak coupling limit (e.g. by substituting Na to K, Rb, or Li). This is particularly so since detailed theoretical analysis \cite{Dirac_semi} shows that one branch of the 3D Dirac cone in Na$_3$Bi is composed of in-plane Bi $p_x, p_y$ orbitals, indicating a fairly weak (quasi-2D) interlayer coupling along $c-$axis. In the very-weak interlayer coupling limit, appropriate doping or gating that moves the chemical potential to the Lifshitz transition energy in Na$_3$Bi may lead to various exotic phenomena such as unconventional superconductivity \cite{BSCCO} or magnetism in the bulk \cite{graphene_Kondo}, or even fractional topological phenomena. Furthermore, interaction can be enhanced by introducing additional symmetry breaking dopings even if the quasi-2D limit is perfected achieved. Another exciting aspect of Na$_3$Bi is that our experimental data provide evidence supporting surface state Fermi arc structures with helical spin polarization. For example, in surface electrical transport experiments at high magnetic field, it is interesting to study how the SS electrons wind around the constant energy arc-contour. Also, if superconductivity can be induced by bulk doping in Na$_3$Bi, then the surface state superconductivity can be topologically nontrivial leading to Majorana modes \cite{Kane_Proximity}, whereas the bulk superconductivity in two 3D Dirac cones may potentially host interesting finite-momentum Cooper-pairing state \cite{Cho}.  On the other hand, breaking time-reversal symmetry (via doping magnetic impurities) or breaking inversion symmetry (certain adiabatic crystal distortion) can split each bulk Dirac node in Na$_3$Bi to two spin polarized Weyl nodes, realizing a topological Weyl semimetal. Furthermore, magnetic doping in Na$_3$Bi that breaks time-reversal symmetry can test whether the surface states are indeed of (Z$_2$) topological origin as a result of the single bulk inversion at $\Gamma$ \cite{Chen Science Fe}.

In summary, we uncover the novel electronic and spin structure in Na$_3$Bi by studying its non-natural-cleavage (100) crystalline surface. We discover the first known bulk material featuring the Lifshitz transition singularity in 3D-Dirac-cone-multiplet electronic structure, additionally with the evidence of 2D nontrivial spin-orbit surface states nestled in between the bulk Dirac cones. The exotic Dirac physics highlighted here can be further tuned in high quality nano-structured samples in future works.

\section{Methods}

\textbf{Sample growth and ARPES measurement techniques:} High quality single crystals of Na$_3$Bi are grown by standard method reported in Ref. \cite{Brauer}. Ultraviolet spin integrated ARPES measurements were performed at Beamline I4 at MAX-lab, Lund, Sweden, and the ESPRESSO endstation installed at Beamline-9B of the Hiroshima Synchrotron Radiation Center (HiSOR), Hiroshima, Japan, using SPECS PHOIBOS 100 and Scienta R4000 electron analyzers respectively. Incident photon energies range from 16 to 80 eV. Spin-resolved ARPES measurements were performed at the ESPRESSO endstation at HiSOR. Photoelectrons are excited by an unpolarized He-I$\alpha$ light (21.21 eV). The spin polarization is detected by state-of-the-art very low energy electron diffraction (VLEED) spin detectors utilizing preoxidized Fe(001)-p($1 \times 1$)-O targets \cite{Okuda_BL9B}. The two spin detectors are placed at an angle of 90$^\circ$ and are directly attached to a VG-Scienta R4000 hemispheric analyzer, enabling simultaneous spin-resolved ARPES measurements for all three spin components as well as high resolution spin integrated ARPES experiments. The energy and momentum resolution was better than 30 meV and $1\%$ of the surface BZ for spin-integrated ARPES measurements at Beamline I4 at MAX-lab, and 80 meV and $3\%$ of the surface BZ for spin-resolved ARPES measurements at ESPRESSO endstation at HiSOR. Samples were cleaved \textit{in situ} under a vacuum condition better than $1 \times 10^{-10}$ torr at both beamlines. Since Na$_3$Bi is extremely air sensitive, argon-filled professional glove boxes with residual oxygen and water level less than 1 ppm are used in the entire preparation process. After preparation, the samples are transferred to the ARPES vacuum chamber via air-tight containers filled with purified argon. The (100) cleavage surface is made possible by carefully choosing sample pieces with flat, (100) surface exposed; multiple pre-cleaving trials are performed to make sure that the crystals are cleaved at the desired surface. Then the actual cleavage plane is checked by the Fermi surface map and the shape of surface BZ in ARPES. For example, one bulk Dirac node per hexagonal surface BZ means that the cleavage is at the (001) surface, whereas two bulk Dirac nodes per rectangular surface BZ means that the cleavage is at the (100) surface. ARPES measurements are performed at liquid nitrogen temperature in BL-I4 at MAX-lab, and 10-20 K in BL9B at HiSOR. Samples are found to be stable and without degradation for a typical measurement period of 8 hours. No charging effect due to insulating behavior was found for all samples measured.

\textbf{First principles calculation methods:} The first-principles calculations are based on the generalized gradient approximation (GGA) \cite{Perdew} using the full-potential projected augmented wave method \cite{Blochl} as implemented in the VASP package \cite{Kress}. Experimental lattice parameters are adopted from Ref. \onlinecite{Na3Bi_crystal}. The electronic structure of bulk Na$_3$Bi are calculated using a $9\times5\times5$ Monkhorst-Pack $k$-mesh over the BZ with the inclusion of spin-orbit coupling. In order to simulate surface effect, we use $14\times1\times1$ and $1\times1\times7$ supercell for two different terminations (100) and (001), respectively, with vacuum thickness larger than 20 $\textrm{\AA}$.  $9\times5$ and $5\times5$ $k$-mesh are used for (100) and (001) termination, respectively.

\newpage

\addtocounter{figure}{-1}
\begin{figure*}[t!]
\caption{\textbf{f}-\textbf{g}, First principles $\bar{K}$-$\bar{\Gamma}$-$\bar{M}$ surface (\textbf{f}) and $\tilde{A}$-$\tilde{\Gamma}$ surface (\textbf{g}) electronic structure. Theoretical nontrivial surface state (green lines) appears along the edge of the bulk 3D Dirac cone in the (001) surface, while it becomes separated from the bulk continuum when seen from the (100) surface (\textbf{g}). \textbf{h}, Overlook of the ARPES band structure from the (100) surface. two separated bulk Dirac nodes (BDP1 and BDP2) are clearly visible at the Fermi level. Linear dispersive pattern of one of the cones is well resolved in the \textit{k}-\textit{E} cuts.}
\end{figure*}

\clearpage

\begin{figure*}
\centering
\includegraphics[width=17cm]{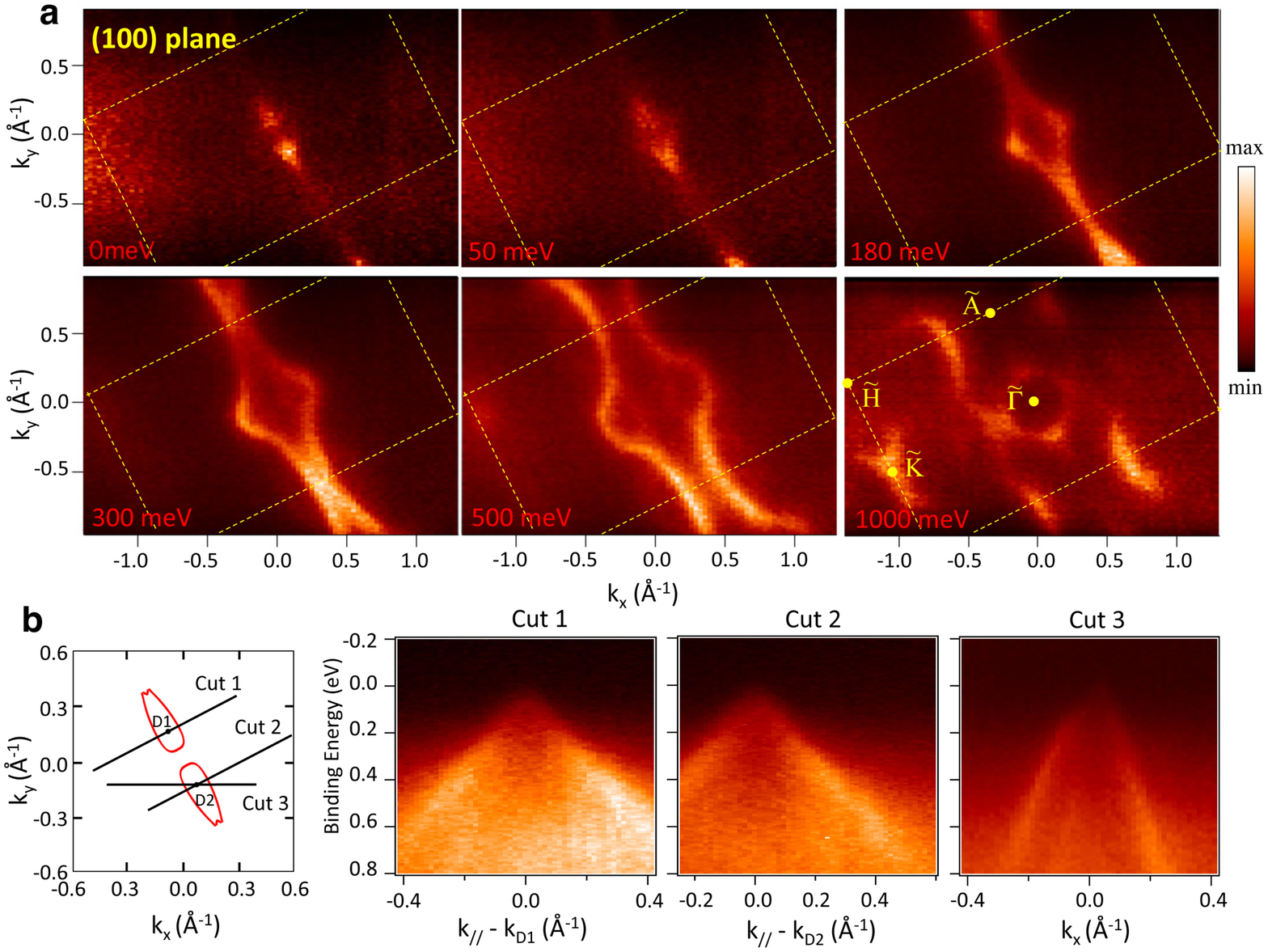}
\caption{\textbf{ARPES observation of bulk Dirac states.} \textbf{a}, ARPES constant energy maps obtained at the (100) surface. Yellow dashed boxes indicate the surface Brillouin zone; notations of high symmetry points are shown at the $E_B = 1000$ meV panel; binding energies are marked in red on the lower left corner of each panel. \textbf{b}, Selected ARPES \textit{k}-\textit{E} cuts, directions and rough $k$-ranges of the cuts are indicated by black solid lines in the left panel.}\label{3Dcones}
\end{figure*}

\newpage

\begin{figure*}
\centering
\includegraphics[width=17cm]{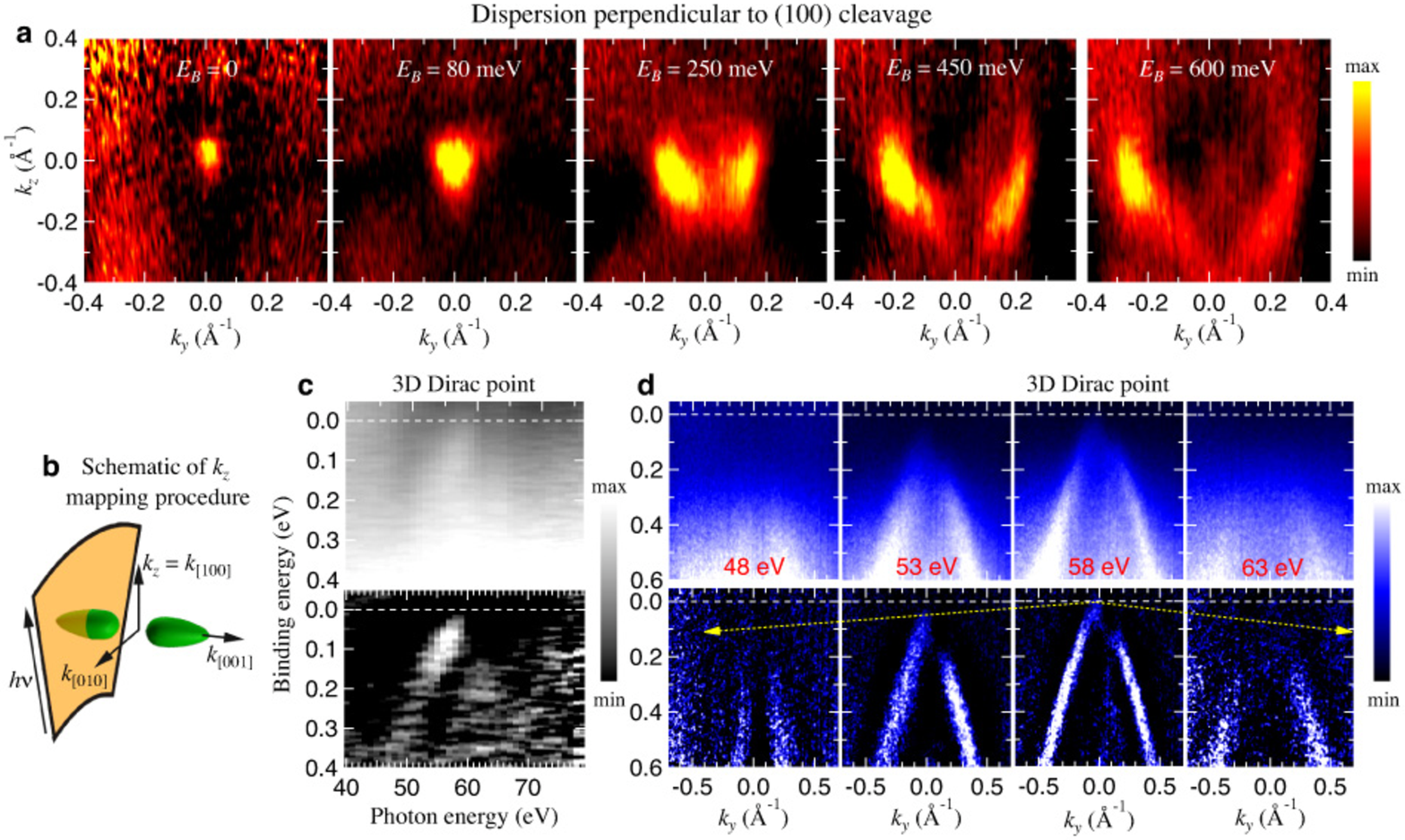}
\caption{\textbf{Out-of-plane dispersion of the bulk Dirac fermions.} \textbf{a}, ARPES constant energy maps along the $k_y$-$k_z$ plane. Binding energies are marked on top of each panel. The $k_z$ [100] axis is built by measuring ARPES $k$-$E$ maps with photon energies ranging from 40 to 76 eV. \textbf{b}, Schematic diagram of the $k_z$ mapping procedure. Since the cleaving surface is (100), $k_z$ corresponds to the [100] direction parallel to the crystallographic $a$ axis. \textbf{c}, Linear dispersion along the third dimension ($k_z$). Probing with different photon energies, the top of one of the bulk Dirac cones locate at different binding energies, constructing the linear $k_z$ dispersion. \textbf{d}, Linear $k_z$ dispersion seen from in-plane $k$-$E$ cuts. It is clear from the data that the bulk Dirac cone does not touch the Fermi level except for $h\nu=58$ eV, establishing again the three dimensional Dirac dispersion. For Panels \textbf{c} and \textbf{d}, top row presents raw data, Bottom row shows second derivative results along the momentum distribution curves.}\label{kz}
\end{figure*}

\newpage

\begin{figure*}
\centering
\includegraphics[width=17cm]{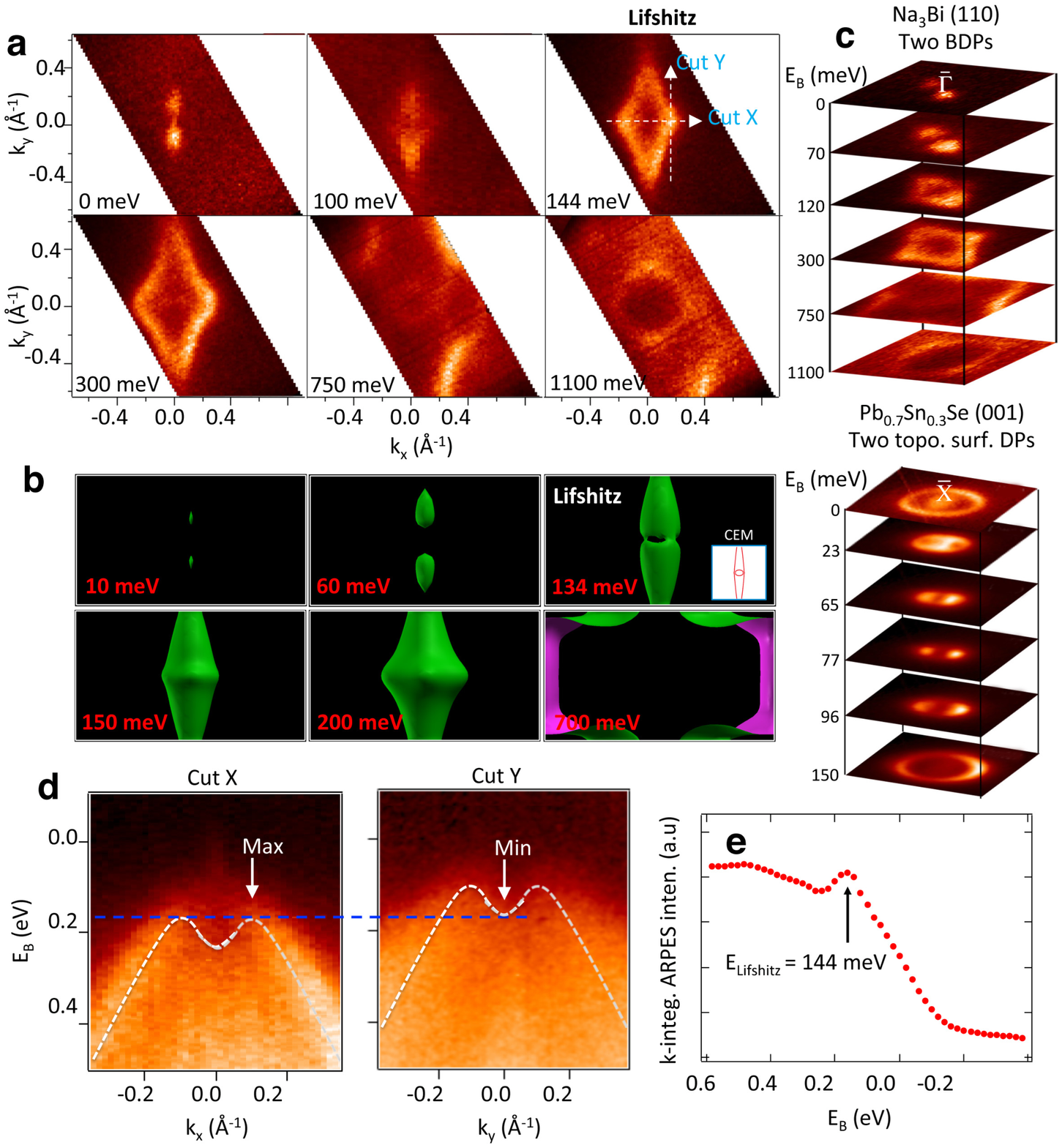}
\caption{\textbf{Lifshitz transition and saddle point singularity in the bulk Dirac cones.} \textbf{a}, Lifshitz transition observed in ARPES constant energy maps. Note that the data set presented here is different from that in Fig. \ref{3Dcones}. At $E_B \sim 144$ meV the two cones touch and hybridize to form a diamond shaped single band contour, marking the Lifshitz transition. \textbf{b}, Theoretical 3D constant energy maps at various binding energies, for comparison with Panel \textbf{a}. Inset in the 134 meV panel shows band projections onto the (100) surface at the Lifshitz energy. CEM: constant energy map. \textbf{c}, Lifshitz transition of \textit{bulk} Dirac fermions in Na$_3$Bi (top panel), }\label{Lifshitz}
\end{figure*}
\addtocounter{figure}{-1}
\begin{figure*}[t!]
\caption{compared with the Lifshitz transition of \textit{surface} Dirac gas in a topological crystalline insulator Pb$_{0.7}$Sn$_{0.3}$Se (bottom panel). \textbf{d}, ARPES $k$-$E$ maps along Cut X and Cut Y in Fig. \ref{Lifshitz}\textbf{a}. The hybridization of the two cones results in a saddle point singularity at the intersection of Cut X and Cut Y, signified by band maximum (lowest binding energy) along $k_x$ (left panel) and band minimum (highest binding energy) along $k_y$. \textbf{e}, Saddle point singularity observed from an anomalous increase of local density of states. The peak of integrated ARPES intensity marks the Lifshitz transition at $E_B \sim 144$ meV.}
\end{figure*}

\begin{figure*}
\centering
\includegraphics[width=17cm]{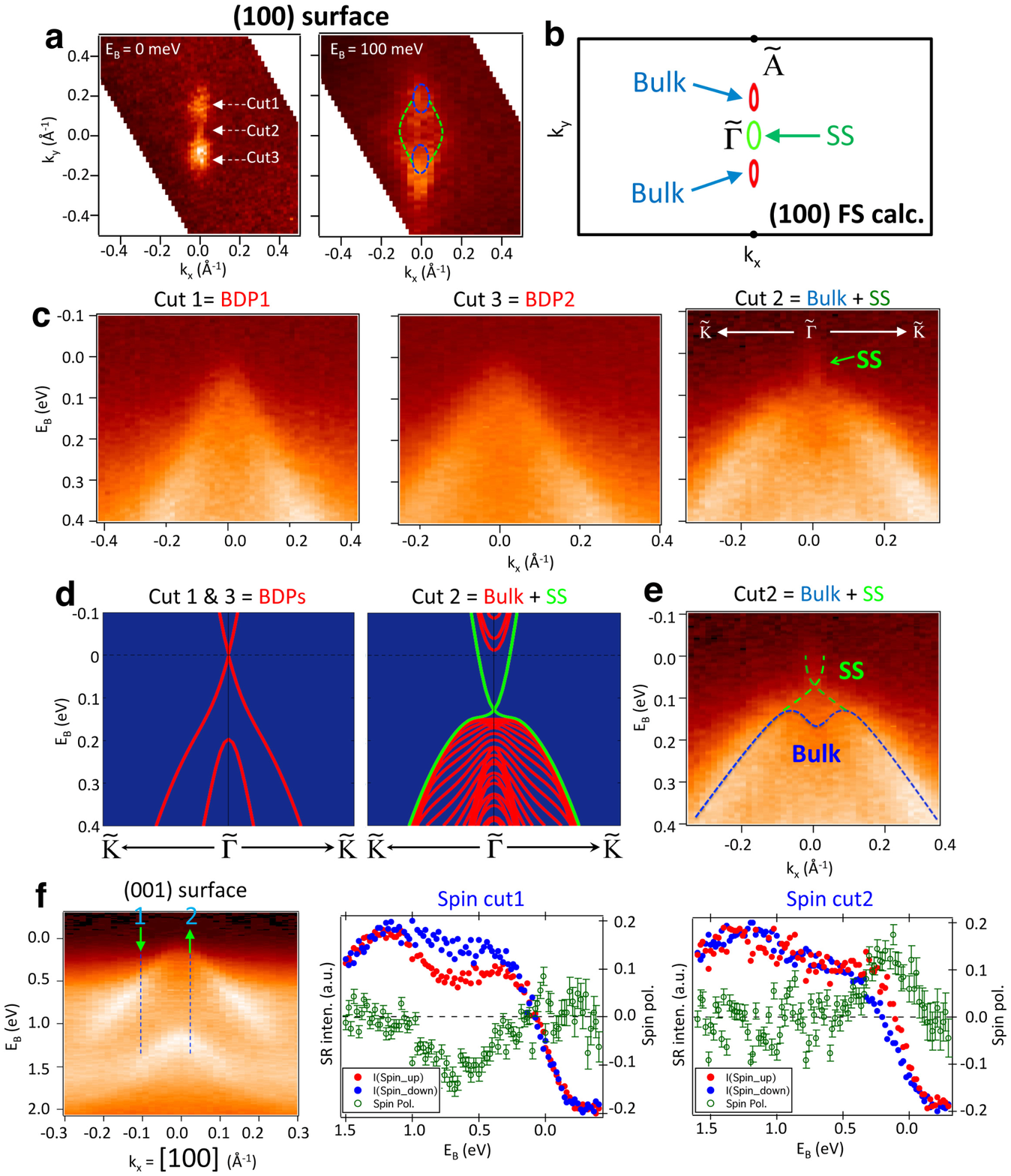}
\caption{\textbf{Signatures for existence of spin-helical nontrivial surface states.} \textbf{a}, Fermi surface (left) and a constant energy map at $E_B = 100$ meV (right) of Na$_3$Bi at the (100) surface. Blue ellipses mark the bulk band contour; green curves denote the possible location of the expanded surface state. The raw data without guides to the eye lines can be found in Fig. 4\textbf{a}. \textbf{b}, Bulk and surface Fermi surface of Na$_3$Bi from first principles slab calculations.}\label{TSS}
\end{figure*}
\addtocounter{figure}{-1}
\begin{figure*}[t!]
\caption{ \textbf{c}, Experimental appearance of the surface state at the (100) surface. Bulk Dirac cones are resolved without trace of surface state at Cuts 1 and 3, while extra ARPES intensity appears at a small $k$-space range at the zone center along Cut 2. \textbf{d}, Results of slab calculation corresponding to Panel \textbf{d}. \textbf{e}, Same $k$-$E$ map as the right panel of \textbf{c} but with bulk and surface states marked by blue and green dashed curves, respectively. \textbf{f}, Experimental spin helical texture of the surface state in another surface termination, the (001) surface. The spin-resolved measurements are performed by using an unpolarized He-I$\alpha$ light (21.21 eV), see method section.}
\end{figure*}

\begin{thebibliography}{99}

\bibitem{Graphene} Geim, A. K., \& Novoselov, K. S. The rise of graphene. \textit{Nat. Mater.} $\mathbf{6}$, 183-191 (2007).

\bibitem{kimgeim}  Geim, A. K., \& Kim, P. Carbon Wonderland. \textit{Sci. Am.} $\textbf{298}$, 90-97 (2008).
 
\bibitem{RMP} Hasan, M. Z. \& Kane, C. L. Topological Insulators. \textit{Rev. Mod. Phys.} $\mathbf{82}$, 3045-3067 (2010).

\bibitem{Zhang_RMP} Qi, X-L. \& Zhang, S-C. Topological insulators and superconductors. \textit{Rev. Mod. Phys.} $\mathbf{83}$, 1057-1110 (2011).


\bibitem{Fang} Fang, Z. \textit{et.al.} The anomalous Hall effect and magnetic monopoles in momentum space. Science $\mathbf{302}$, 92-95 (2003).
\bibitem{Ashvin} Wan, X. \textit{et al.} Topological semimetal and Fermi-arc surface states in the electronic structure of pyrochlore iridates. \textit{Phys. Rev. B} $\mathbf{83}$, 205101 (2011).

\bibitem{Balent} Halasz, G. B., \& Balents, L. Time-reversal invariant realization of the Weyl semimetal phase. \textit{Phys. Rev. B} $\mathbf{85}$, 035103 (2012).



\bibitem{3D_Dirac} Murakami, S. Phase transition between the quantum spin Hall and insulator phases in 3D: emergence of a topological gapless phase. \textit{New J. Phys.} $\mathbf{9}$, 356 (2007).


\bibitem{Dirac_3D} Young, S. M. \textit{et al.} Dirac semimetal in three dimensions. \textit{Phys. Rev. Lett.} $\mathbf{108}$, 140405 (2012).

\bibitem{Dirac_semi} Wang, Z. \textit{et al.} Dirac semimetal and topological phase transitions in A$_3$Bi (A = Na, K, Rb). \textit{Phys. Rev. B} $\mathbf{85}$, 195320 (2012).

\bibitem{Dai} Wang, Z. \textit{et.al.} Three dimensional Dirac semimetal and quantum spin Hall effect in Cd$_3$As$_2$. \textit{Phys. Rev. B} $\mathbf{88}$, 125427 (2013).

\bibitem{Yu Science QAH} Yu, R. \textit{et al}. Quantized anomalous Hall effect in magnetic topological insulators. \textit{Science} $\mathbf{329}$, 61-64 (2010).


\bibitem{Zhang Axion} Li, R., Wang, J., Qi X.-L. \& Zhang S.-C., Dynamical axion field in topological magnetic insulators. \textit{Nat. Phys.} \textbf{6}, 284-288 (2010).

\bibitem{Kim} Dean, R. C. \textit{et al}. Hofstadter's butterfly in moire superlattices: A fractal quantum Hall effect. \textit{Nature} $\mathbf{497}$, 598-602 (2013).

\bibitem{Geim} Ponomarenko, L. A. \textit{et al}. Cloning of Dirac fermions in graphene superlattices. \textit{Nature} doi:10.1038/nature12187 (2013).

\bibitem{Weyl superconductor} Meng, T., \& Balents, L. Weyl superconductors. \textit{Phys. Rev. B} $\mathbf{86}$, 054504 (2012).



\bibitem{Suyang} Xu, S.-Y. \textit{et al.} Topological phase transition and texture inversion in a tunable topological insulator. \textit{Science} $\mathbf{332}$, 560-564 (2011).


\bibitem{Oh} Brahlek, M. \textit{et al.} Topological-metal to band-insulator transition in (Bi$_{1-x}$In$_x$)$_2$Se$_3$ thin films. \textit{Phys. Rev. Lett.} $\mathbf{109}$, 186403 (2012).

\bibitem{Note} Throughout this paper, we use $(nml)$ to denote the crystalline surface, [$nml$] to denote crystalline direction/axis. For example, at the (001) surface, or along the [001] direction.

\bibitem{CdAs_Hasan} Neupane, M. \textit{et.al.} Observation of a topological 3D Dirac semimetal phase in high-mobility Cd$_3$As$_2$. Preprint at http://arXiv:1309.7892 (2013).

\bibitem{CdAs_Cava} Borisenko, S. \textit{et.al.} Experimental realization of a three-dimensional Dirac semimetal. Preprint at http://arXiv:1309.7978 (2013).

\bibitem{Chen_Na3Bi} Liu, Z.~K. \textit{et al.}, Experimental realization of a three-dimensional Dirac semimetal, Na$_3$Bi. Preprint at http://arXiv:1310.0391 (2013).


\bibitem{Na3Bi_crystal} Massalski, T. B. Binary alloy phase diagrams (ASM, Materials Park, 1990).

\bibitem{Helical_metal} Bergman, D. L. \& Refael, G. Bulk metals with helical surface states. \textit{Phys. Rev. B} $\mathbf{82}$, 195417 (2010).


\bibitem{Okada} Okada, Y. \textit{et al.} Observation of Dirac node formation and mass acquisition in a topological crystalline insulator. \textit{Science} $\mathbf{341}$, 1496-1499 (2013).


\bibitem{Kane_Proximity} Fu, L., \& Kane, C. L. Superconducting Proximity Effect and Majorana Fermions at the Surface of a Topological Insulator. \textit{Phys. Rev. Lett.} $\mathbf{100}$, 096407 (2008).

\bibitem{Quasi-2D}Hirsch. J.E., Scalapino. D. J., Enhanced Superconductivity in Quasi Two-Dimensional Systems. \textit{Phys. Rev. Lett.} $\mathbf{56}$, 2732-2735 (1986).

\bibitem{BSCCO} King. D. M. \textit{et al}. Observation of a Saddle-Point Singularity in Bi$_2$(Sr$_{0.97}$Pr$_{0.03}$)$_2$CuO$_{6+\delta}$ and Its Implications for Normal and Superconducting State Properties. \textit{Phys. Rev. Lett.} $\mathbf{73}$, 3298 (1994).
\bibitem{Pnictide} Liu. C. \textit{et al}. Evidence for a Lifshitz transition in electron-doped iron arsenic superconductors at the onset of superconductivity. \textit{Nature Phys.} $\mathbf{6}$, 419 (2010).

\bibitem{graphene_Kondo}Lipinski, S., Krychowski, D.  Kondo effect near the van Hove singularity in biased bilayer graphene. 	arXiv:1206.4455 (2013)


\bibitem{Cho} Cho, G. Y.  Bardarson, J. H.  Lu, Y.-M., Moore, J. E. Superconductivity of doped Weyl semimetals: finite-momentum pairing and electronic analogues of the 3He-A phase. \textit{Phys. Rev. B} $\mathbf{86}$, 214514 (2012).

\bibitem{Chen Science Fe}  Chen, Y.-L. \textit{et al}. Massive Dirac Fermion on the Surface of a Magnetically Doped Topological Insulator. \textit{Science} $\mathbf{329}$, 659-662 (2010).

\bibitem{Brauer}
Brauer, G. \& Zintl, E. Konstitution von Phosphiden, Arsoniden, Antimoniden und Wismutiden des Lithiums, Natriums und Kaliums. \textit{Z. Phys. Chem., Abt. B} $\mathbf{37}$, 323-352 (1937).

\bibitem{Okuda_BL9B}
Okuda, T. \textit{et al.} Efficient spin resolved spectroscopy observation machine at Hiroshima Synchrotron Radiation Center. \textit{Rev. Sci. Instrum.} {\bf 82}, 103302 (2011).

\bibitem{Perdew} Perdew, J. P., Burke, K., \& M. Ernzerhof, Generalized Gradient Approximation Made Simple. \textit{Phys. Rev. Lett.} $\mathbf{77}$, 3865 (1996).
\bibitem{Blochl} Bl\"ochl, P. E. Projector augmented-wave method. \textit{Phys. Rev. B.} $\mathbf{50}$, 17953 (1994).

\bibitem{Kress} Kress G., \& Hafner, J. $Ab$ $initio$ molecular dynamics for open-shell transition metals. \textit{Phys. Rev. B.} $\mathbf{48}$, 13115 (1993).
\end{thebibliography}
\end{document}